\documentclass{icrc2009}

\usepackage{graphicx}   
\usepackage[caption=false]{caption}    
\usepackage[font=footnotesize]{subfig} 
\usepackage{fixltx2e}
\usepackage{url}

\newcommand{\shorttitle}[1]%
{\markboth{Proceedings of the 31\MakeLowercase{$^{st}$} ICRC, {\L}\'{o}d\'{z} 2009}{#1} }
\newcommand{\etal}{\MakeLowercase{\textit{et al. }}} 

\begin{document}
\title{Multiwavelength Observations of Mrk 501 in 2008}

\author{\IEEEauthorblockN{D. Kranich\IEEEauthorrefmark{1},
    D. Paneque\IEEEauthorrefmark{6},
    A. Cesarini\IEEEauthorrefmark{14},
    A. Falcone\IEEEauthorrefmark{2},
    M. Giroletti\IEEEauthorrefmark{3},
    E. Hoversten\IEEEauthorrefmark{2},
    T. Hovatta\IEEEauthorrefmark{4}, \\
    Y.Y. Kovalev\IEEEauthorrefmark{5},
    A. L\"ahteenm\"aki\IEEEauthorrefmark{4},
    E. Nieppola\IEEEauthorrefmark{4},
    C. Pagani\IEEEauthorrefmark{2},
    A. Pichel\IEEEauthorrefmark{7},
    K. Satalecka\IEEEauthorrefmark{8}, \\
    J. Scargle\IEEEauthorrefmark{9},
    D. Steele\IEEEauthorrefmark{10},
    F. Tavecchio\IEEEauthorrefmark{11},
    D. Tescaro\IEEEauthorrefmark{12}, 
    M. Tornikoski\IEEEauthorrefmark{4} and
    M. Villata\IEEEauthorrefmark{13} \\
    on behalf of the MAGIC and VERITAS collaborations$^0$}
  \\
  \IEEEauthorblockA{\IEEEauthorrefmark{1} ETH Zurich, CH-8093 Switzerland}
  \IEEEauthorblockA{\IEEEauthorrefmark{6} SLAC National Accelerator
    Laboratory and KIPAC, CA 94025, USA}
  \IEEEauthorblockA{\IEEEauthorrefmark{2} Penn State University,
    Astronomy \& Astrophysics Dept., University Park,  PA 16802, USA}
  \IEEEauthorblockA{\IEEEauthorrefmark{3} INAF Istituto di
    Radioastronomia, Bologna, Italy}
  \IEEEauthorblockA{\IEEEauthorrefmark{4} Mets\"ahovi Radio Observatory,
    Helsinki University of Technology TKK, Finland}
  \IEEEauthorblockA{\IEEEauthorrefmark{5} MPIfR, 53121 Bonn, Germany
    and ASC Lebedev, 117997 Moscow, Russia}
  \IEEEauthorblockA{\IEEEauthorrefmark{7} Instituto de Astronomia y
    Fisica del Espacio Ciudad Universitaria, Buenos Aires, Argentina}
  \IEEEauthorblockA{\IEEEauthorrefmark{8} DESY Deutsches
    Elektronen-Synchrotron, D-15738 Zeuthen, Germany}
  \IEEEauthorblockA{\IEEEauthorrefmark{9} NASA Ames Research Center,
    Moffett Field, CA 94035, USA}
  \IEEEauthorblockA{\IEEEauthorrefmark{10} Adler Planetarium \&
    Astronomy Museum, Chicago, IL 60605, USA}
  \IEEEauthorblockA{\IEEEauthorrefmark{11} INAF National Institute for
    Astrophysics, I-00136 Rome, Italy}
  \IEEEauthorblockA{\IEEEauthorrefmark{12} IFAE, Campus UAB, E-08193
    Bellaterra, Spain}
  \IEEEauthorblockA{\IEEEauthorrefmark{13} INAF Osservatorio
    Astronomico di Torino, Italy}
  \IEEEauthorblockA{\IEEEauthorrefmark{14}  Centre for Astronomy, Physics Department, National University of Ireland, Galway, Ireland.}}

\shorttitle{D. Kranich \etal Mrk 501 MWL campaign 2008}
\maketitle

\begin{abstract}
The well-studied VHE ($E >100\ \mathrm{GeV}$) blazar Mrk 501 was
observed between March and May 2008 as part of an extensive
multiwavelength observation campaign including  radio, optical, X-ray
and VHE gamma-ray instruments. Mrk 501 was in a low state of activity
during the campaign, with a low VHE flux of about 20\% the Crab Nebula
flux. Nevertheless, significant flux variations could be observed in
X-rays as well as $\gamma$-rays. Overall Mrk 501 showed increased
variability when going from radio to $\gamma$-ray energies.\\
The broadband spectral energy distribution during the two different
emission states of the campaign was well described by a homogeneous
one-zone synchrotron self-Compton model. The high emission state was
satisfactorily modeled by increasing the amount of high energy
electrons with respect to the low emission state. This
parameterization is consistent with the energy-dependent variability
trend observed during the campaign.
\end{abstract}


\footnotetext{For a full author list, please see R. Ong et al (these
  proceedings) or
  http://veritas.sao.arizona.edu/conferences/authors?icrc2009"
  (VERITAS) and http://wwwmagic.mpp.mpg.de/collaboration/members/
  (MAGIC)}

\section{Introduction}

Mrk 501 is a well-studied nearby (redshift $z=0.034$) blazar which was
first detected at TeV energies by the Whipple collaboration in 1996
\cite{Quinn1996}. In subsequent years Mrk 501 was regularly observed
and detected in VHE $\gamma$-rays by many other Cherenkov telescope
experiments. In particular during the whole year 1997 when it showed
an exceptionally strong outburst with peak flux levels up to 10 times
the Crab Nebula flux and flux-doubling time scales down to 0.5 days
\cite{Aharonian1999}. Mrk 501 also showed strong flaring activity at
X-ray energies during that year. The X-ray spectrum obtained was very
hard and the synchrotron peak was found to be at $\sim100\
\mathrm{keV}$, about 2 orders of magnitude higher than in previous
observations \cite{Pian1998}. In the following years, Mrk 501 showed
only low $\gamma$-ray emission (of the order of 20-30\% the Crab
Nebula flux), apart from a few single flares of higher intensity. In
2005, the MAGIC telescope was able to observe Mrk 501 during another
high-emission state which, although at a lower flux level compared to
1997, showed flux variations of an order of magnitude and
unprecedented flux doubling time scales (down to a few minutes)
\cite{Albert2007}. Mrk 501 has been the target of many multiwavelength
(MWL) campaigns (e.g. \cite{Kataoka1999, Krawczynski2000,
  Tavecchio2001, Anderhub2009}), mainly covering the object during
flaring activity.\\
The data presented here were taken between March 25th and May 16th,
2008 during an extended MWL campaign covering radio (Effelsberg, IRAM,
Medicina, Mets\"ahovi, Noto, RATAN-600, VLBA), optical (KVA), UV
(Swift/UVOT), X-ray (RXTE/PCA, Swift/XRT and Swift/BAT) and
$\gamma$-ray (MAGIC, Whipple, VERITAS) energies. The duration as well
as the energy coverage of this particular Mrk 501 campaign are rather
unique. Details on the participating instruments and the data analysis will
be presented in an upcoming paper \cite{FutureM501Paper}.

\section{Light Curves}

 \begin{figure*}[th]
  \centering
  \includegraphics[width=4.8in]{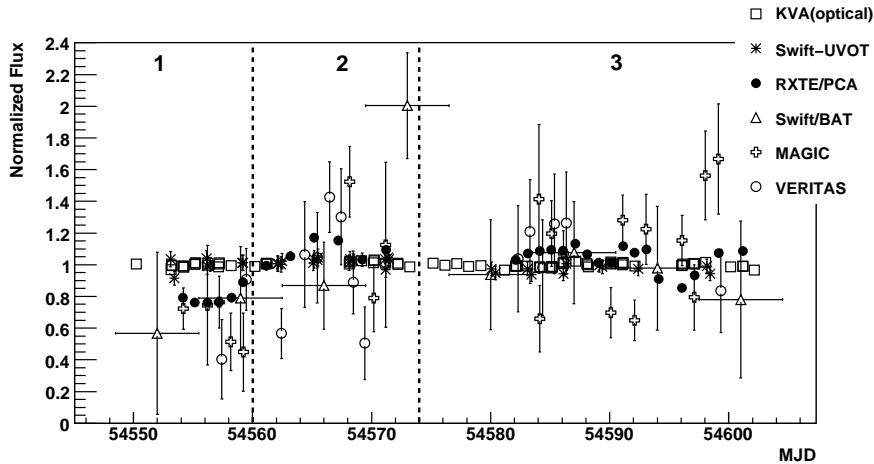}
  \caption{Combined normalized light curves for a selection of the
    instruments taking part in the campaign. The vertical bars denote
    $1\ \sigma$  statistical uncertainties, and the horizontal bars
    the integration time of the observation.}
  \label{fig_lightcurve}
 \end{figure*}

Figure \ref{fig_lightcurve} shows the normalized light
curves\footnote{Each individual light curve was normalized to its
  average.} for a selection of the instruments involved in the
campaign. The different light curves cover the optical $R$ band (KVA),
the UV band (Swift/UVOT), the soft  X-ray band (RXTE/PCA), the hard
X-ray band (Swift/BAT) and the VHE band (MAGIC and VERITAS). The
average fluxes for each instrument are given as: $4.4\ \mathrm{mJy}$
(KVA), \mbox{$1.6\ \mathrm{mJy}$} (Swift/UVOT), \mbox{$2.8\cdot10^{-4}
  \mathrm{counts/s}$} (Swift/BAT), \mbox{$8.2\cdot10^{-11}
  \mathrm{erg/cm^2/s}$} (RXTE/PCA), \mbox{$2.7\cdot10^{-11}
  \mathrm{ph/cm^2/s}$} (MAGIC) and \mbox{$2.2\cdot10^{-11}
  \mathrm{ph/cm^2/s}$} (VERITAS). Other instruments providing valuable
data (like Swift/XRT or Whipple) have been omitted for the sake of
clarity in this plot. Flux variations are large in X-rays and
$\gamma$-rays, but rather small in the UV and optical. Due to the
small error bars in the X-ray data, the most significant flux
variations can be observed at these energies. The plot also shows some
evidence for a correlated flux variability at X-rays and VHE
$\gamma$-rays (see section \ref{sec:cross}) indicating a low-emission
state before MJD 54560 and a somewhat stronger emission
afterwards. For the spectral analysis presented below we divided the
data set into three time intervals taking into account the X-ray flux
level (i.e. low/high flux before/after MJD 54560) and the data gap at
most frequencies around MJD 54574.

\section{Variability}

We followed the description given in \cite{Vaughan2003} to quantify
the flux variability by means of the so-called fractional variability
parameter $F_{\mathrm{var}}$. In order to account for the individual
flux measurement errors ($\sigma_{\mathrm{err, i}}$), the `excess
variance' (\cite{Nandra1997, Edelson2002}) was used as an estimator of
the intrinsic source flux variance. This is the variance after
subtracting the contribution expected from measurement
errors. $F_{\mathrm{var}}$ was derived for each individual instrument
taking part in the campaign, which covered an energy range from radio
frequencies at $\sim$8 GHz up to very high energies at $\sim$10
TeV. $F_{\mathrm{var}}$ is calculated as:

\begin{equation}
  F_{\mathrm{var}} = \sqrt{\frac{S^2 - <\sigma_{\mathrm{err}}^2
      >}{<F_{\gamma}>^2}}
  \label{form_nva}
\end{equation}
where  $<F_{\gamma}>$ denotes the average photon flux, $S$ the
standard deviation of the $N$ flux measurements and
\mbox{$<\sigma_{\mathrm{err}}^2>$} the mean squared error, all
determined for a given instrument (energy bin). The uncertainty of
$F_{\mathrm{var}}$ is estimated according to:

\begin{equation}
\Delta F_{\mathrm{var}} = \frac{<\sigma_{\mathrm{err}}^2>}{\sqrt{N}
  <F_{\gamma}>} \cdot \sqrt{1 + \frac{1}{2<F_{\gamma}>^2
    F_{\mathrm{var}}^2}} \nonumber
\end{equation}

\begin{figure}[!t]
  \centering
  \includegraphics[width=3.in]{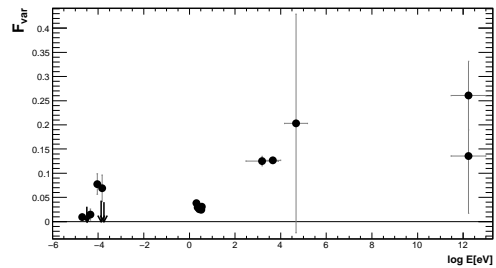}
  \caption{Fractional variability parameter $F_{\mathrm{var}}$ for all
    the instruments participating in the campaign. $F_{\mathrm{var}}$
    was derived using the individual single-night flux measurements
    except for Swift/BAT for which data integrated over one week were
    used. Vertical bars denote $1\ \sigma$ uncertainties, horizontal
    bars indicate the approximate energy range covered by the
    instrument. The arrows indicate 95\% confidence level upper
    limits.}
  \label{fig_fracvar}
\end{figure}

Fig. \ref{fig_fracvar} shows the $F_{\mathrm{var}}$ values derived for
all instruments that participated in the MWL campaign. Some
instruments showed a negative excess variance
($<\sigma_{\mathrm{err}}^2>$ larger than $S^2$), which can happen when
there is little variability and/or the errors are slightly
overestimated. Essentially such a result can be interpreted as no
signature for variability in the data of that particular instrument,
either because a) there was no variability or b) the instrument was
not sensitive enough to detect it. In these cases, upper limits of
95\% confidence level were computed.\\
The plot, on the other hand, also shows significant variability
detected with various other instruments during the
campaign. Essentially all instruments observing at optical or larger
frequencies recorded variability. The plot also shows some evidence
that the recorded flux variability increases with energy: in the
optical $R$ band (ground-based telescopes) and the 6 filters from
Swift/UVOT the variability is around 2-4\%, in X-rays it is about
13\%, and at VHE at the 20\% level, although affected by large error
bars (due to the large uncertainties in the flux measurements). The
radio instruments show no evidence for variability, with the exception
of RATAN (22 GHz) and Mets\"ahovi (37 GHz) that show $\sim
~7\pm2\%$.\\
In the synchrotron self-Compton (SSC) framework, the observed flux
variability contains information on the dynamics of the underlying
population of relativistic electrons (and possibly positrons). In this
context, the general variability trend reported in
Fig. \ref{fig_fracvar} suggests that  the flux variations are produced
by the injection of energetic particles, which are characterized by
shorter cooling time scales, causing the higher variability amplitude
observed at the highest energies.

\section{Multifrequency cross-correlations \label{sec:cross} }
\begin{figure}[!t]
  \centering
  \subfloat[RXTE
  vs. MAGIC \& VERITAS]{\includegraphics[width=2.6in]{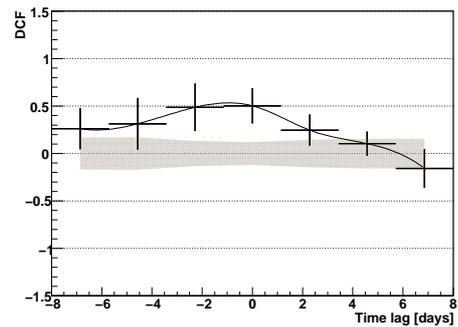}
    \label{fig_dcf1}} \hfil
  \subfloat[RXTE
  vs. XRT]{\includegraphics[width=2.6in]{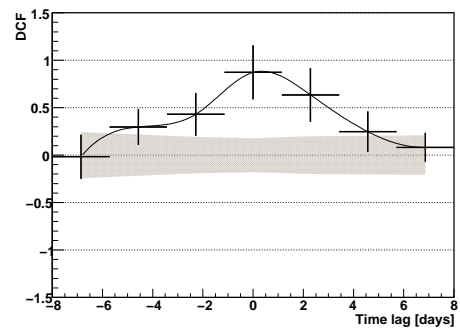}
    \label{fig_dcf2}}
  \caption{Discrete Correlation Function for time lags from -8 to +8
    days in steps of $\sim2$ days. The grey band represents the
    expected fluctuation of the DCF values in the case of completely
    uncorrelated time series given the error bars from the actual
    observations.}
\end{figure}

In order to study the multifrequency cross-correlations between the
different energy bands we used the Discrete Correlation Function (DCF)
as described  in \cite{Edelson1988}. This method can also be applied
in the case of unevenly sampled data as taken in this campaign.\\
The DCF was derived for all different combinations of instruments /
energy regions and also for artificially introduced time lags (ranging
from -8 to +8 days) between the individual light curves. Such time
lags may occur as a result of spatially separated emission regions of
the individual flux components (as expected, for example, in external
inverse Compton models), or may be caused by the energy dependent
cooling time-scales of the emitting electrons.\\
Based on the MWL data from this campaign, significant correlations
have been found for the pairs RXTE/PCA - Swift/XRT and also (less
significant) RXTE/PCA (or Swift/XRT) with MAGIC and VERITAS
(Fig. \ref{fig_dcf1} and \ref{fig_dcf2}). In both cases, the DCF
maximum is obtained for a zero time lag with a value of $0.87\pm0.28$
(RXTE/PCA - Swift/XRT ) and $0.5\pm0.19$ (RXTE/PCA - MAGIC and
VERITAS) respectively. Due to the modest flux variability and / or large
flux errors, no strong conclusions could be drawn from this analysis.

\section{SED modeling}

\begin{figure*}[th]
  \centering
  \includegraphics[width=4in]{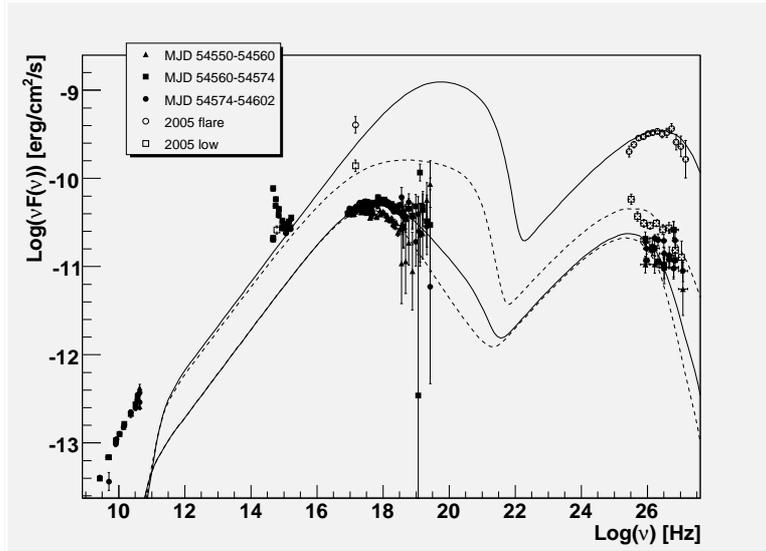}
  \caption{Broadband SED for Mrk 501 as obtained during this campaign
    in comparison to low and high states from 2005. The results from a
    SSC model fit to the low state 2008 data (dot-dashed cuve), the
    high state 2008 data (heavy-dashed curve), the 2005 low state
    (light-dashed curve) and the 2005 high state (solid line curve)
    are shown.}
  \label{fig_sed}
\end{figure*}

\begin{table}[!h]
  \caption{The SSC model parameters used to describe the broadband
    SED for different flux states of the campaign.}
  \label{table_modelpars}
  \centering
  \begin{tabular}{|c|c|c|c|c|}
  \hline
     &  2008 & 2008 & 2005 & 2005\\
     &  high state & low state & high & low\\
   \hline
    &&&&\\
    $\gamma_{\mathrm{break}}$ & $2.6 \cdot 10^5$  & $2.2 \cdot 10^5$ &
    $1.0 \cdot 10^6$  & $1.0 \cdot 10^5$\\ 
    n$_1$ & 2.0 & 2.0 & 2.0 & 2.0\\
    n$_2$ & 3.9 & 4.2 & 3.9 & 3.2\\
    B [G] & 0.19 & 0.19 & 0.23 & 0.31 \\
    K [cm$^{-3}]$ & $1.8 \cdot 10^4$ & $1.8 \cdot 10^4$ & $7.5 \cdot
    10^4$ & $4.3 \cdot 10^4$ \\
    R [cm] & $3 \cdot 10^{15}$ & $3 \cdot 10^{15}$ & $1 \cdot 10^{15}$
    & $1 \cdot 10^{15}$ \\
    $\delta$ & 12 & 12 & 25 & 25\\
  \hline
  \end{tabular}
  \end{table}

The broadband SED of Mrk 501 for the three different time periods
defined above, together with some historical data from the 2005 low
and high state of the object are shown in Fig. \ref{fig_sed}. The
host galaxy contribution ($12.0 \pm 0.3\ \mathrm{mJy}$
\cite{Nilsson2007}) has been subtracted from the optical (KVA) data
while the $\gamma$-ray spectra have been corrected for EBL absorption
using the `low-IR' model of \cite{Kneiske2004}. The results from a
one-zone SSC model fit to the different data sets are also shown in the
figure as dashed lines. The model code was developed by Tavecchio et 
al. (\cite{Tavecchio1998, Tavecchio2001}) and is based on the
following characteristic parameters: a spherical emission region with
radius $R$ and Doppler factor $\delta$, a magnetic field of strength
$B$, an electron distribution (density $K$) following a broken power
law with slopes $n_1$ and $n_2$ and break energy
$\gamma_{\mathrm{break}}$. The actual values of these model parameters
for the two different emission states during the campaign and the
historical data from 2005 are given in Tab. \ref{table_modelpars}. As
can be seen from Fig. \ref{fig_sed} the model is able to accurately
reproduce the data at X-ray energies. Given the relatively small
differences in the SEDs of the two emission states of the campaign,
only marginal changes of the model parameters were required in order
to adjust the model to the two states. The proposed explanation for
the low - high state transition is the injection of fresh, high-energy
electrons which lead to a shift of the $\gamma_{\mathrm{break}}$
energy and to a hardening of the spectrum.\\
The discrepancy between the model and the data at lower energies
(radio, optical) can be caused by synchrotron radiation from
additional, cooler electron populations which could be present at 
different locations in the jet. The higher (than expected) fluxes at
radio/optical frequencies  were discussed in the past (also with Mrk501
data) in the framework of  the helical-jets in blazar scenarios
\cite{Villata1999} or the blob-in-jet scenario
\cite{Katarzynski2001}. As is shown in table \ref{table_modelpars},
in comparison to the historical 2005 SED, the model parameters have
changed significantly. However, it is worth noticing that the sparse
coverage of the 2005 data allow for a lot of degeneracy among the
(large) number of model parameters. A robust statement from the
comparison of the 2005 and 2008 SEDs is that, while the X-ray and
gamma-ray fluxes did change substantially between these two epochs,
the fluxes at optical frequencies remained approximately the same. In
the framework of two populations of electrons, this result suggests
that the population of cool electrons does not vary with time while
the population of electrons responsible for the X-ray (Synchrotron)
and gamma-ray (Inverse Compton) emission is very dynamic. A more detailed modeling of the 
experimental data will be performed in a forthcoming publication \cite{FutureM501Paper}.

\section* {ACKNOWLEDGMENTS}
We thank the Instituto de Astrofisica de Canarias for the excellent
working conditions at the Observatorio del Roque de los Muchachos in
La Palma. The support of the German BMBF and MPG, the Italian INFN and
Spanish MCINN is gratefully acknowledged. This work was also supported
by ETH Research Grant TH 34/043, by the Polish MNiSzW Grant N N203
390834, and by the YIP of the Helmholtz Gemeinschaft.\\
This research is supported by grants from the US Department of Energy,
the US National Science Foundation, and the Smithsonian Institution, 
by NSERC in Canada, by Science Foundation Ireland, and by STFC in the UK.
We acknowledge the excellent work of the technical support staff at the FLWO
and the collaborating institutions in the construction and operation of
VERITAS.\\
Y.Y.K. is a research fellow of the Alexander von~Humboldt Foundation.


\begin{thebibliography}{99}

 \bibitem{Quinn1996} J.~Quinn, \etal 1996, ApJ, 456, L831
 \bibitem{Aharonian1999} F.~Aharonian, \etal 1999, A\&A, 342, 69
 \bibitem{Pian1998} E.~Pian, \etal 1998, ApJ, 492, L17
 \bibitem{Albert2007} J.~Albert, \etal 2007, ApJ, 669, 862
 \bibitem{Kataoka1999} J.~Kataoka, \etal 1999, ApJ, 514, 138
 \bibitem{Krawczynski2000} H.~Krawczynski, \etal 2000, A\&A, 353, 97
 \bibitem{Tavecchio2001} F.~Tavecchio, \etal 2001, ApJ, 554, 725
 \bibitem{Anderhub2009} H.~Anderhub, \etal 2009, submitted to ApJ
 \bibitem{FutureM501Paper} D. Paneque \etal 2009, SLAC-PUB-13628, in preparation.
 \bibitem{Vaughan2003} S.~Vaughan, \etal 2003, MNRAS, 345, 1271
 \bibitem{Nandra1997} K.~Nandra, \etal 1997, ApJ, 476, 70
 \bibitem{Edelson2002} R.~Edelson, \etal 2002, ApJ, 568, 610
 \bibitem{Edelson1988} R.~Edelson and J.~Krolik. 1988, ApJ, 333, 646
 \bibitem{Nilsson2007} K.~Nilsson, \etal 2007, A\&A, 475, 199
 \bibitem{Kneiske2004} T.~Kneiske, \etal 2004, A\&A, 413, 807
 \bibitem{Tavecchio1998} F.~Tavecchio, \etal 1998, ApJ, 509, 608
\bibitem{Villata1999} M.~Villata and C.M.~Raiteri 1999, A\&A, 347,30
\bibitem{Katarzynski2001} K.~Katarzynski, \etal 2001, A\&A, 367,809


 \end{thebibliography}
\end{document}